\documentclass[iop]{emulateapj}
\usepackage{amsmath}
\usepackage[flushleft]{threeparttable}
\usepackage{array}
\usepackage{float}
\usepackage{graphicx}
\usepackage{subfigure}
\usepackage{natbib}
\usepackage{color}
\usepackage{dcolumn}
\newcolumntype{d}[1]{D{.}{.}{#1}}

\begin{document}
\title{$0.7-2.5~\mu$\MakeLowercase{m} spectra of Hilda asteroids}
\author{Ian Wong,\altaffilmark{1} Michael E. Brown,\altaffilmark{1} and Joshua P. Emery\altaffilmark{2}}
\affil{\textsuperscript{1}Division of Geological and Planetary Sciences, California Institute of Technology, Pasadena, CA 91125, USA; iwong@caltech.edu \\
\textsuperscript{2}Earth and Planetary Science Department \& Planetary Geosciences Institute, University of Tennessee, Knoxville, TN 37996, USA}

\begin{abstract}
The Hilda asteroids are primitive bodies in resonance with Jupiter whose origin and physical properties are not well understood. Current models posit that these asteroids formed in the outer Solar System and were scattered along with the Jupiter Trojans into their present-day positions during a chaotic episode of dynamical restructuring. In order to explore the surface composition of these enigmatic objects in comparison with an analogous study of Trojans \citep{emery}, we present new near-infrared spectra (0.7--2.5~$\mu$m) of 25 Hilda asteroids. No discernible absorption features are apparent in the data. Synthesizing the bimodalities in optical color and infrared reflectivity reported in previous studies, we classify 26 of the 28 Hildas in our spectral sample into the so-called less-red and red sub-populations and find that the two sub-populations have distinct average spectral shapes. Combining our results with visible spectra, we find that Trojans and Hildas possess similar overall spectral shapes, suggesting that the two minor body populations share a common progenitor population. A more detailed examination reveals that while the red Trojans and Hildas have nearly identical spectra, less-red Hildas are systematically bluer in the visible and redder in the near-infrared than less-red Trojans, indicating a putative broad, shallow absorption feature between 0.5 and 1.0~$\mu$m. We argue that the less-red and red objects found in both Hildas and Trojans represent two distinct surface chemistries and attribute the small discrepancy between less-red Hildas and Trojans to the difference in surface temperatures between the two regions. 
\end{abstract}

\section{Introduction}

The Hilda asteroids are a large population of minor bodies located in the 3:2 mean-motion resonance with Jupiter. These objects orbit within a relatively narrow range of heliocentric distances around 4.0~AU, between the outer edge of the Main Belt (roughly 3.3~AU) and Jupiter's orbit (5.2~AU). and Early theories of solar system formation posited that the Hildas originated in the middle Solar System and were captured into their present-day orbits during a period of smooth migration \citep[e.g.,][]{franklin}. However, recent advances in our understanding of solar system evolution have placed Hildas in a new light. Current models describe a scenario in which the gas giants crossed a mutual mean-motion resonance sometime after the era of planet formation, triggering chaotic alterations throughout the Solar System \citep[e.g.,][]{morbidelli}. These models predict that a significant fraction of planetesimals that formed beyond the primordial orbits of the ice giants was scattered inward during the period of dynamical instability and now resides in the middle Solar System \citep{gomes}.

In particular, simulations carried out within this dynamical framework have demonstrated that the present-day Hildas and Jupiter Trojan, both resonant asteroid populations in the vicinity of Jupiter, should be comprised almost exclusively of objects from the outer Solar System \citep{roig2}. Such a scenario also points toward a common progenitor population for the Kuiper Belt objects, Trojans, and Hildas, presenting the enticing possibility that studying the surface properties of the more readily accessible Hildas could give insight into the composition of all three populations. The compositional characterization of Hildas and comparison with the properties of Trojans serve as a powerful observational test for probing the veracity of current dynamical instability models.

At present, the composition of Hildas, and indeed that of all other middle and outer solar system minor bodies, remains a subject of much speculation. The general compositional outline for these asteroids posits a mixture of water ice, anhydrous silicates, and organics \citep[e.g.,][]{vilas,emerybrown,yangjewitt}. Observational studies of Hildas have hitherto revealed largely featureless, reddish spectra in the optical and near-infrared \citep{Hildaspectra,Hildaspectra2,takiremery} and very low albedos averaging around 4\% \citep{Hildaalbedo}. Taxonomically, Hildas are predominantly D- and P-type asteroids, with a small fraction of C-type objects \citep{grav}. Further into the infrared, a handful of published Hilda spectra reveal a broad, rounded feature centered at around 3~$\mu$m, which has been interpreted to be due to a thin layer of water frost coating a dark-grained regolith \citep{takiremery}. A study of a similar feature on Trojans shows that this feature could also be consistent with ammonia ice or irradiation products thereof \citep{brown}; such a surface chemistry would naturally point toward an outer solar system origin. 

Analysis of optical colors derived from Sloan Digital Sky Survey photometry and infrared reflectivities measured by the Wide-field Infrared Survey Explorer reveals a strong bimodality among both Hildas \citep{gilhutton,wong4} and Trojans \citep[e.g.,][]{roig,wong}. This result demonstrates that Hildas and Trojans are comprised of two spectrally distinct sub-populations and signifies a new dimension in the study of these objects. Comparative spectroscopy promises to expand our understanding of the surface composition, the underlying cause(s) of the sub-populations' bimodal features, possible divergent evolutionary signatures between Hildas and Trojans, and ultimately, the origin of these objects in the broader context of theories of solar system formation and evolution.

\citet{emery} analyzed near-infrared spectra of 68 Trojans and uncovered a highly robust bifurcation in spectral slope that corresponds with the previously described bimodality in optical colors. In this paper, we describe the results from an analogous near-infrared spectroscopic survey of Hildas. These observations were undertaken with the objective of compiling a significant body of high-quality Hilda spectra to enable detailed comparison with the earlier results of \citet{emery}, as well as continuing the search for spectral features that may help further constrain the surface composition of these poorly-understood objects. 

\section{Observations and Data Reduction}\label{sec:obs}

\begin{table*}[t!]
	\centering
	\begin{threeparttable}
		\caption{Observation Details} \label{tab:obs}	
		\renewcommand{\arraystretch}{1.2}
		\begin{center}
			\begin{tabular}{ l  c  l  c  c  d{2.1} c  c  l  c  c  c } 
				\hline\hline
				Number &  Name &  \multicolumn{1}{c}{Date} &  \multicolumn{1}{c}{Start Time} &  $t_{\mathrm{int}}$ &  \multicolumn{1}{c}{$V$\textsuperscript{a}}  & \multicolumn{1}{c}{$H_{v}$}  &\multicolumn{1}{c}{Diam\textsuperscript{b}}& \multicolumn{1}{c}{Standard} & Spectral  & $V-J$ & $J-K$ \\
				 &   &  \multicolumn{1}{c}{(UT)} &  \multicolumn{1}{c}{(UT)}  &  (min) &   &  & \multicolumn{1}{c}{(km)} & \multicolumn{1}{c}{Star} & Type &  & \\
				\hline
				499 &  Venusia &  2016 Mar 12 &  14:40 &  20  & 16.4 & 9.39  & 88  & HD 133011 & G2V & 1.16 & 0.33  \\
				748 &  Simeisa &  2016 Mar 12 &  15:11 &  20  & 16.0 & 9.01  & 105  & HD 140990 & G2V & 1.13 & 0.34 \\
				1162 &  Larissa &  2016 Apr 2 &  06:23 &  20  & 16.2 & 9.44  & 86  & HD 60298 & G2V & 1.23 &  0.34 \\
				1180 &  Rita &  2016 Apr 2 &  05:35 &  32  & 16.5 & 9.14  & 99  & HD 259516 &  G2V & 1.13 & 0.34 \\
				1256 &  Normannia &  2016 Aug 5 &  07:41 &  8  & 15.4 & 9.66  & 78 & HD 172404 & G2V & 1.09 & 0.32 \\
				1268 &  Libya &  2016 Mar 12 &  12:54 &  12  & 15.4 & 9.12  & 100  & HD 129829 &  G2V & 1.03 &  0.32 \\
				1269 &  Rollandia &  2016 Aug 5 &  14:40 &  24  & 15.6 & 8.82  & 114  & HD 10861 & G2V  & 1.09 & 0.34 \\
				1512 &  Oulu &  2016 Apr 2 &  06:52 &  20  & 16.2 & 9.62  & 79  & HD 77730 & G2V & 1.18 &  0.39 \\
				1578 &  Kirkwood &  2016 Mar 15 &  13:49 &  92  & 17.2 & 10.26  & 59  & HD 136983 & G1V & 1.14 & 0.34 \\
				1746 &  Brouwer (1) &  2016 Mar 12 &  15:53 &  12  & 17.0 & 9.95  & 68  & HD 132412 & G2V & 1.09 & 0.31 \\
				1746 &  Brouwer (2) &  2016 Mar 15 &  12:20 &  60  & 16.9 & 9.95  & 68  & HD 132412 & G2V & 1.09 & 0.31   \\
				1754 &  Cunningham &  2016 Aug 5 &  09:42 &  8  & 14.7 & 9.77  & 74  & HD 197089 & G2V & 1.07 & 0.35 \\
				1902 &  Shaposhnikov &  2016 Apr 16 &  05:17&  60  & 16.9 & 9.51  & 83  & HD 259516 &  G2V & 1.13 & 0.34 \\
				1911 &  Schubart (1) &  2016 Mar 12 &  12:28 &  16  & 16.0 & 10.11  & 63  & HD 116367 &  G3V & 1.15 &  0.38 \\
				1911 &  Schubart (2) &  2016 Mar 15 &  11:58 &  12  & 15.9 & 10.11  & 63  & HD 116367 &  G3V & 1.15 &  0.38 \\
				2067 &  Aksnes &  2016 Apr 5 &  12:17 &  64  & 17.1 & 10.48  & 53  & HD 139485 & G5V & 1.32 & 0.46 \\
				2312 &  Duboshin &  2016 Mar 12 &  13:15 &  60  & 17.0 & 10.18  & 61  & HD 124071 & G1V & 1.09 &  0.34\\
				2624 &  Samitchell &  2016 Aug 5 &  08:47 &  20  & 16.0 &10.8  & 46  & HD 177518 & G2V & 1.29 & 0.34 \\
				2760 & Kacha (1) &  2016 Mar 12 &  12:05 &  12  & 15.6 &10.04  & 65  & HD 114962 & G1V & 1.14 & 0.29 \\
				2760 & Kacha (2) &  2016 Mar 15 &  11:43 &  8  & 15.6 & 10.04  & 65  & HD 114962 &  G1V & 1.14 & 0.29 \\
				3561 & Devine &  2016 Aug 5 &  09:57 &  28  & 16.0 &11.1  & 40  & HD 196164 & G2V & 1.10 &  0.31\\
				3577 & Putilin &  2016 Aug 5 &  13:02 &  76  & 16.7 &10.4  & 55  & HD 224251 & G2V & 1.20 &  0.38 \\
				4446 & Carolyn &  2016 Aug 5 &  09:22 &  12  & 15.2 & 11.2  & 38  & HD 197089 &  G2V & 1.07 & 0.35  \\
				5603 & Rausudake (1) &  2016 Mar 12 &  11:29 &  16  & 16.0 & 10.7  & 48  & HD 98503 & G5V & 1.16 & 0.37 \\
				5603 & Rausudake (2) &  2016 Mar 15 &  11:13 &  20  & 16.0 & 10.7  & 48  & HD 98503 &  G5V & 1.16 & 0.37 \\
				5661 & Hildebrand &  2016 Aug 5 &  12:36 &  16  & 15.2 & 11.1  & 40  & HD 203311 & G2V & 1.12 & 0.36 \\
				9121 & Stefanovalentini &  2016 Aug 5 &  10:54 &  72  & 16.4 &10.7  & 48  & HD 196164 &  G2V & 1.10 &  0.31 \\
				11388 & 1998 VU4 &  2016 Apr 16 &  07:19 &  88  & 17.2 &11.2  & 38  & HD 78536 & G3V & 1.12 & 0.41 \\
				14669 & Beletic (1) &  2016 Apr 5 &  11:28 &  36  & 17.4 & 11.5  & 33  & HD 120204 & G2V & 1.08 & 0.31 \\
				14669 & Beletic (2) &  2016 Apr 16 &  09:24 &  20  & 17.2 & 11.5  & 33  & HD 117860 & G2V & 1.13 & 0.36 \\
				 	\hline		
			\end{tabular}
			
			\begin{tablenotes}
				\small
				\item {\bf Notes.} 
				\item \textsuperscript{a}Visible apparent magnitudes at the time of observation.
				\item \textsuperscript{b}Diameters are calculated from the listed absolute magnitudes ($H_{v}$) assuming $p_{v}=0.04$ \citep{Hildaalbedo}.
								
			\end{tablenotes}
		\end{center}
	\end{threeparttable}
\end{table*}

We carried out five observing runs at the NASA Infrared Telescope Facility (IRTF) throughout 2016. The spectra were obtained with the medium-resolution spectrograph and imager SpeX \citep{rayner}. We used the LoRes prism mode with a $0.8\times 15$~arcsec slit, which produces single-order spectra spanning the wavelength range 0.7--2.5~microns. A total of 25 Hildas were observed, of which 5 were observed on two nights.

Our observing strategy closely mirrored the methods described in \citet{emery}. Objects were dithered 7.5~arcsec along the slit between pairs of observations. In order to minimize readout time, while reducing the effect of atmospheric variability (in particular, the OH emission at these wavelengths), single-exposure integration times were set at 120~s for all of our observations. The telescope tracked each object using the automatic guider, which measures the spillover of the object outside of the slit and corrects the pointing to center the object in the slit at a rate of several times a minute. Solar analog G-dwarfs within $5^{\circ}$ of the asteroid were regularly observed --- typically every $\sim$30 minutes, or whenever the airmass of the object changed by 0.10--0.15. The effect of differential refraction across the wavelength range was minimized by aligning the slit with the parallactic angle ($\pm20^{\circ}$) for all asteroid and star observations. Flat-field and argon lamp wavelength calibration frames were taken at the beginning or end of each observing run. Details of our Hilda observations are listed in Table~\ref{tab:obs}.

The prism data were reduced following standard procedures for near-infrared spectra. We utilized the IDL-based spectral reduction program Spextool \citep{cushing} in our data reduction. For each pair of dithered frames, this program handles flat-field removal, wavelength calibration, non-linearity correction, background subtraction, and spectral extraction through a graphical user interface. To correct for telluric absorptions as well as the solar spectrum, each extracted asteroid spectrum was divided by the corresponding calibration star spectrum that was obtained closest in time and airmass. The effects of instrument flexure on the wavelength calibration and telluric correction were addressed by shifting each asteroid spectrum relative to the corresponding calibration star spectrum to minimize the variability within the water absorption regions. Bad pixels and other significant outliers were manually pruned from the extracted spectra in Spextool before the individual dithered-pair spectra for each object were combined into a single reflectance spectrum.

\section{Results and Discussion}

The normalized reflectance spectra of the 25 Hildas targeted in our observing runs are shown in the Appendix. Here, and in the subsequent analysis, we have included additional spectra of three Hildas (153 Hilda, 190 Ismene, and 361 Bononia) obtained using IRTF/SpeX and published in \citet{takiremery}. The signal-to-noise ratio (S/N) of the 28 Hilda spectra in $K$-band (2.22~$\mu$m) range from 20 to 300.

As is the case with the 0.7--2.5~micron Trojan spectra published in \citet{emery}, none of the Hilda spectra show any absorption features to within the noise level in the data. Therefore, while the surface composition of Hildas is typically assumed to be similar to that of Trojans, i.e., rich in water ice, organics, and crystalline silicates \citep{emerybrown,yangjewitt}, no evidence for these materials are found in the near-infrared spectra.

\subsection{Classification into Sub-populations}\label{subsec:class}
Several previous studies have revealed that Hildas are bimodal with respect to visible color and infrared reflectivity. Using these results, we can classify objects into sub-populations and examine their average near-infrared spectra separately.

Analysis of photometry for Hildas contained in the Sloan Digital Sky Survey Moving Object Catalog (SDSS-MOC) demonstrates that the distribution of spectral slopes at visible wavelengths is strongly bimodal, indicative of two sub-populations with visible colors centered at $4.0$ and $9.3$, in units of $\%/1000~\mathrm{\AA}$ \citep{gilhutton,wong4}. Following previous works, we refer to these two sub-populations as less-red (LR) and red (R). Ten of the 28 Hildas in our sample have SDSS-MOC photometry; an additional eight objects have earlier published visible spectral slopes \citep{Hildaspectra,Hildaspectra2}. We note that there is significant overlap between the two modes in the visible color distribution, so some objects with intermediate spectral slopes cannot be definitively categorized as LR or R. The sub-population classifications based on visible colors are listed in Table~\ref{tab:colors}. We classify 8 objects as LR, 6 objects as R, and 4 objects as intermediate.

The reflectivity of Hildas in the W1 (3.4~$\mu$m) and W2 (4.6~$\mu$m) bands of the Wide-field Infrared Survey Explorer (WISE) space telescope also allows for classification of Hildas into the LR and R sub-populations. Specifically, the distribution of relative infrared reflectance in the W1/W2 bands with respect to visible albedo shows two clearly separated clusters \citep[see Figure~12 in][]{grav}, with one group systematically more reflective at infrared wavelengths than the other. Selecting for Hildas with reflectance measurements in both W1 and W2 bands, we are able to categorize 18 objects (9 LR and 9 R). The sub-population classifications based on infrared reflectivity are listed in Table~\ref{tab:colors}.

Comparing the classifications for objects with both visible spectral slope and infrared reflectivity data, we see that the classifications from these two independent methods are consistent, i.e., objects categorized as LR via visible spectral slope are also categorized as LR based on infrared reflectivity, and likewise for R objects. This same pattern was observed for Trojans \citep{wong} and indicates that LR and R objects differ systematically and predictably in both the visible and the infrared. Combining the two sets of classifications, we are able to categorize 22 out of 28 Hildas in our near-infrared spectral sample: 11 LR and 11 R. We can classify an additional 4 objects (3 R and 1 LR) in our spectral sample, which do not have independent published optical or infrared observations, based on their relative positions in near-infrared color space (see Section~\ref{subsec:colors}). This brings the total classification count to 12 LR and 14 R.

\begin{figure}[t!]
\begin{center}
\includegraphics[width=9cm]{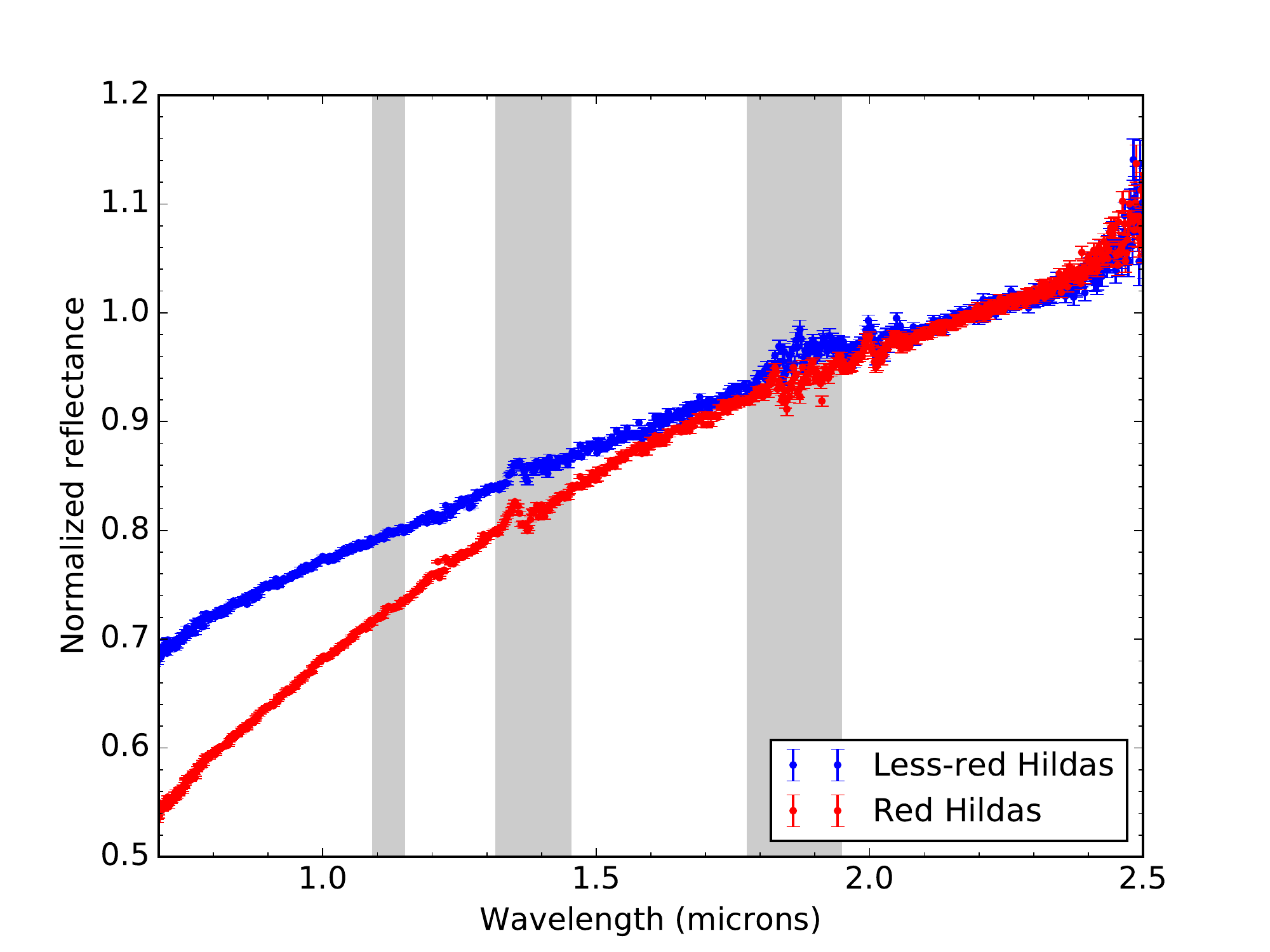}
\end{center}
\caption{Average of spectra in the LR and R Hilda sub-populations, normalized to unity at 2.2~$\mu$m. The two spectra are averages of 12 and 14 individual object spectra, respectively. Gray bars mark regions of strong telluric water vapor absorption. The two sub-populations are distinguished primarily by the difference in spectral slope at shorter near-infrared wavelengths ($\lambda < 1.5$). The error bars shown here and in subsequent figures are the uncertainties on the individual reflectance values derived from the weighted average of individual object spectra.} \label{average}
\end{figure}

Figure~\ref{average} plots the average of all spectra in the two color sub-populations. The main observation is that the average LR and R Hilda spectra are highly distinct, with the difference in shape concentrated primarily at the shorter end of the near-infrared range ($\lambda < 1.5$). While the R Hilda spectrum is concave down throughout most of the near-infrared, the LR Hilda spectrum is mostly straight, with only a slight downturn at the shortest wavelengths. This indicates that the LR and R Hilda sub-populations have distinct surface properties, in agreement with the bimodality previously seen at both visible and infrared wavelengths.

\subsection{Comparison with Jupiter Trojans}\label{subsec:trojans}

In order to compare the absolute reflectivity spectra of LR and R Hildas and Trojans, we compiled all previously published visible spectra \citep{bus,lazzaro,takiremery} as well as broadband SDSS-MOC photometry of Hildas within our near-infrared spectral sample.\footnote{We do not include the visible spectra from \citet{Hildaspectra} and \citet{Hildaspectra2}, as those spectra are not published with uncertainties on the individual data points.} All five Hildas with visible spectra (153 Hilda, 190 Ismene, 361 Bononia, 1180 Rita, and 1754 Cunningham) are LR objects. The visible spectral and photometric data are matched to the near-infrared reflectance spectra at 0.75 and 0.913 (Sloan \textit{z}-band)~$\mu$m, respectively. The absolute reflectivity level is set by the average visible (0.55~$\mu$m) geometric albedo: 0.04 \citep{Hildaalbedo}. 

\begin{figure}[t!]
\begin{center}
\includegraphics[width=9cm]{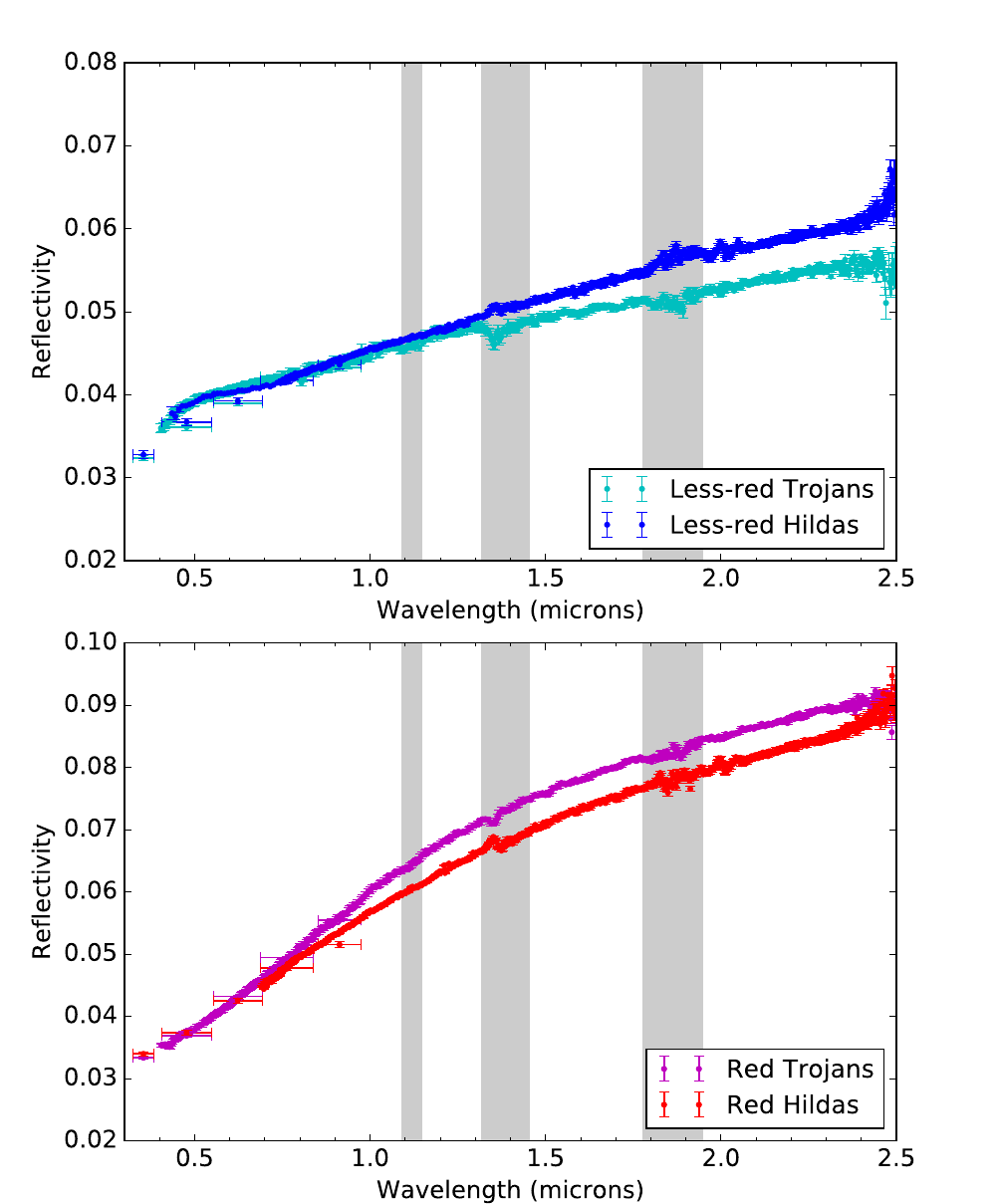}
\end{center}
\caption{Comparison of combined visible and near-infrared average spectra of LR (top) and R (bottom) Hildas and Trojans. The absolute reflectivity level at 0.55~$\mu$m for each spectrum is set by the published visible albedos: 0.04 for Hildas \citep{Hildaalbedo} and 0.041 for Trojans \citep{fernandez}. Average $u-g-r-i-z$ broadband reflectances for objects in our spectral sample with SDSS-MOC photometry are overplotted. R Hildas and Trojans have largely identical spectra. On the other hand, LR Hildas are systematically bluer in the visible and redder in the near-infrared than LR Trojans, suggesting the presence of a broad, shallow absorption feature between 0.5 and 1.0~$\mu$m.} \label{trojans}
\end{figure}

Figure~\ref{trojans} shows the average reflectivity spectra of LR and R Hildas, along with analogous data for Trojans taken from \citet{emery} and \citet{takiremery}. We note that the Hilda and Trojan visible spectral samples are a subset of the corresponding near-infrared spectral samples; moreover, there is little overlap between the visible spectral and SDSS-MOC photometric samples. Therefore, some mismatch between the visible spectral and photometric data can be expected.

The main observation from the comparison plot is that the corresponding sub-populations in the Hildas and Trojans have notably similar spectra. Looking at the R objects separately, we find that R Hildas and R Trojans display largely identical spectral slopes across the visible and near-infrared wavelengths. This observation is corroborated by the reported average visible spectral slopes: 9.3~$\%/1000~\mathrm{\AA}$ for R Hildas \citep{wong4}, and 9.6~$\%/1000~\mathrm{\AA}$ for R Trojans \citep{wong}. A slight difference in spectral slope is apparent between 0.7 and 0.9~$\mu$m. While this disparity between the two populations may be real, this region corresponds to the lower end of the IRTF/SpeX wavelength range, where the transmission function rises steeply. As such, data in this wavelength region are especially susceptible to residual signals from telluric correction, as, for example, in the case of uncorrected nonlinear effects from instrument flexure (Section~\ref{sec:obs}).

The average LR Hilda and Trojan spectra likewise demonstrate similar overall shapes. However, a closer look reveals the possible presence of a very broad and shallow absorption feature in the region 0.5--1.0~$\mu$m. We note that the sharp inflection point at 0.7~$\mu$m is likely to be primarily caused by the concatenation of different data sets for the visible and near-infrared average spectra, with the former being a subset of the latter. We note that none of the individual Hilda spectra spanning both visible and near-infrared wavelengths shows a discernible feature in this wavelength region above the level of the uncertainties.

Nevertheless, both visual comparison and color indices (see below) demonstrate that LR Hildas are noticeably redder on average (i.e., have steeper spectral slopes) than LR Trojans throughout much of the near-infrared (0.7--2.0~$\mu$m).  In contrast, at visible wavelengths, LR Hildas and somewhat bluer than LR Trojans, as evidenced by their average spectral slopes: 4.0~$\%/1000~\mathrm{\AA}$ for LR Hildas \citep{wong4}, and 5.3~$\%/1000~\mathrm{\AA}$ for LR Trojans \citep{wong}. Together, these disparities imply an overall concave-up shape in the average LR Hilda spectrum relative to the LR Trojan spectrum extending from the visible into the near-infrared. The implications of such an absorption feature are discussed in Section~\ref{subsec:implications}.

\begin{table*}[t!]
	\centering
	\begin{threeparttable}
		\caption{Near-infrared Color Indices and Sub-population Classifications} \label{tab:colors}	
		\renewcommand{\arraystretch}{1.2}
		\begin{center}
			\begin{tabular}{ l  c  c c c c c c c c  } 
				\hline\hline
				Number &  Name &  $0.85-J$ & $0.85-H$ & $0.85-K$ & $J-H$ & $J-K$ & $H-K$ & Vis\textsuperscript{a} & IR\textsuperscript{a} \\
				 &   &   &  &  &  &  &  & Class. & Class. \\
				\hline
				153\textsuperscript{b} &  Hilda & $0.107\pm 0.007$ & $0.202\pm 0.008$ & $0.314\pm 0.008$& $0.096\pm0.009$& $0.208\pm0.009$& $0.112\pm0.010$ & LR & LR \\
				190\textsuperscript{b} &  Ismene & $0.098\pm0.006$ & $0.173\pm0.008$& $0.282\pm0.009$& $0.075\pm0.008$&$0.184\pm0.010$ & $0.109\pm0.011$ & LR & LR \\
				361\textsuperscript{b} &  Bononia & $0.199\pm0.013$ & $0.330\pm0.013$& $0.476\pm0.013$& $0.130\pm0.012$& $0.276\pm0.012$& $0.146\pm0.012$ & LR & LR \\
				499 &  Venusia & $0.077\pm0.009$ & $0.170\pm0.011$ & $0.280\pm0.012$& $0.093\pm0.012$& $0.203\pm0.013$& $0.111\pm0.014$ & LR & LR  \\
				748 &  Simeisa & $0.167\pm0.012$ &$0.292\pm0.012$ & $0.410\pm0.012$& $0.125\pm0.013$& $0.242\pm0.013$& $0.117\pm0.013$& I & LR \\
				1162 &  Larissa &$0.175\pm0.010$  &$0.300\pm0.007$ & $0.413\pm0.011$& $0.126\pm0.011$& $0.238\pm0.014$& $0.112\pm0.012$& -- & R \\
				1180 &  Rita &  $0.157\pm0.012$& $0.270\pm0.011$& $0.380\pm0.012$& $0.113\pm0.013$& $0.224\pm0.013$& $0.111\pm0.012$& LR & LR \\
				1256 &  Normannia &  $0.276\pm0.019$& $0.459\pm0.016$& $0.585\pm0.016$& $0.183\pm0.018$& $0.309\pm0.018$& $0.126\pm0.015$& R & R\\
				1268 &  Libya & $0.179\pm0.013$ & $0.329\pm0.015$& $0.467\pm0.012$& $0.150\pm0.015$& $0.289\pm0.012$& $0.138\pm0.014$& I & LR \\
				1269 &  Rollandia & $0.272\pm0.018$ &$0.427\pm0.017$ & $0.551\pm0.016$& $0.154\pm0.016$& $0.279\pm0.014$& $0.124\pm0.012$& R & R \\
				1512 &  Oulu & $0.214\pm0.015$ & $0.326\pm0.015$& $0.455\pm0.015$& $0.112\pm0.014$& $0.241\pm0.014$& $0.130\pm0.013$&I & R \\
				1578 &  Kirkwood & $0.236\pm0.017$ &$0.371\pm0.014$ & $0.484\pm0.015$& $0.134\pm0.015$& $0.248\pm0.016$& $0.113\pm0.013$& R& R\\
				1746 &  Brouwer & $0.270\pm0.019$ & $0.460\pm0.017$& $0.576\pm0.017$& $0.190\pm0.017$& $0.306\pm0.016$& $0.116\pm0.013$& -- & R \\
				1754 &  Cunningham & $0.134\pm0.011$ &$0.235\pm0.010$ &$0.333\pm0.010$ &$0.101\pm0.011$ & $0.199\pm0.010$& $0.098\pm0.010$& LR & -- \\
				1902 &  Shaposhnikov & $0.121\pm0.010$ & $0.232\pm0.013$& $0.362\pm0.015$& $0.110\pm 0.014$& $0.241\pm0.016$& $0.130\pm 0.018$& -- & LR \\
				1911 &  Schubart & $0.059\pm 0.006$ & $0.128\pm 0.008$& $0.224\pm 0.009$& $0.069\pm 0.008$& $0.164\pm 0.009$& $0.095\pm0.010$& LR& LR\\
				2067 &  Aksnes & $0.066\pm 0.005$ &$0.123\pm 0.006$ &$0.235\pm0.014$ & $0.057\pm 0.006$& $0.169\pm 0.014$& $0.113\pm 0.014$& -- & -- \\
				2312 &  Duboshin & $0.312\pm 0.022$ & $0.484\pm 0.020$& $0.617\pm 0.019$& $0.172\pm 0.019$& $0.306\pm 0.018$& $0.133\pm 0.016$& -- & R\\
				2624 &  Samitchell & $0.194\pm 0.013$ &$0.292\pm 0.015$ & $0.377\pm 0.012$& $0.098\pm 0.014$& $0.183\pm0.011$& $0.085\pm 0.013$& R & -- \\
				2760 & Kacha & $0.211\pm0.014$ &$0.386\pm0.014$ & $0.522\pm 0.012$& $0.175\pm 0.016$& $0.311\pm 0.013$& $0.136\pm0.014$& -- & -- \\
				3561 & Devine & $0.280\pm0.019$ &$0.453\pm0.017$ & $0.580\pm0.017$& $0.174\pm0.016$& $0.300\pm0.016$& $0.127\pm0.014$& I & --  \\
				3577 & Putilin & $0.250\pm0.019$ & $0.382\pm0.016$& $0.515\pm0.017$& $0.133\pm 0.015$& $0.265\pm 0.016$& $0.132\pm 0.013$& R & R \\
				4446 & Carolyn & $0.279\pm0.019$ & $0.431\pm 0.017$& $0.538\pm0.017$& $0.153\pm0.014$& $0.259\pm 0.014$& $0.106\pm0.011$& R & -- \\
				5603 & Rausudake & $0.251\pm0.016$ & $0.431\pm0.014$& $0.592\pm0.014$& $0.180\pm0.016$& $0.341\pm0.016$& $0.161\pm0.014$& --  & R \\
				5661 & Hildebrand & $0.248\pm0.016$ & $0.410\pm0.017$& $0.523\pm0.018$& $0.161\pm0.014$& $0.275\pm0.015$& $0.113\pm0.016$& -- & -- \\
				9121 & Stefanovalentini & $0.183\pm0.013$ &$0.313\pm0.011$ & $0.401\pm0.014$& $0.129\pm0.013$& $0.218\pm0.015$& $0.088\pm0.014$& -- & -- \\
				11388 & 1998 VU4 & $0.013\pm0.010$ & $0.047\pm0.013$& $0.101\pm0.018$& $0.034\pm0.014$& $0.088\pm0.019$& $0.054\pm0.021$& LR & -- \\
				14669 & Beletic & $0.269\pm0.014$  & $0.402\pm0.012$& $0.533\pm0.015$& $0.133\pm0.014$& $0.264\pm0.017$& $0.131\pm0.015$& -- & -- \\
				
				  				\hline
				
			\end{tabular}
			
			\begin{tablenotes}
				\small
				\item {\bf Notes.} 
				\item \textsuperscript{a}Classification into less-red (LR) and red (R) sub-populations based on visible spectral slopes \citep{Hildaspectra,Hildaspectra2,gilhutton,wong4} and infrared albedos measured by WISE \citep{grav}. An intermediate (I) classification denotes a published spectral slope measurement that does not allow for categorization of the object as less-red or red.
				\item \textsuperscript{b}Derived from spectra published in \citet{takiremery}.
								
			\end{tablenotes}
		\end{center}
	\end{threeparttable}
\end{table*}

\subsection{Near-infrared Colors}\label{subsec:colors}

To study the spectra more quantitatively, we calculated near-infrared colors. Following the methods described in \citet{emery}, we derived color indices from the normalized reflectance at four wavelengths: 0.85~$\mu$m, $J$ (1.25~$\mu$m), $H$ (1.65~$\mu$m), and $K$ (2.22~$\mu$m). The color $m_{\lambda1}-m_{\lambda2}$ corresponding to a given reflectance ratio $R_{\lambda2}/R_{\lambda1}$ was calculated using the relation $m_{\lambda1}-m_{\lambda2} = 2.5\log(R_{\lambda2}/R_{\lambda1})$. Standard error propagation was used to calculate the color uncertainties. The measured colors for all the Hildas in our sample are listed in Table~\ref{tab:colors}.

\begin{figure}[t!]
\begin{center}
\includegraphics[width=9cm]{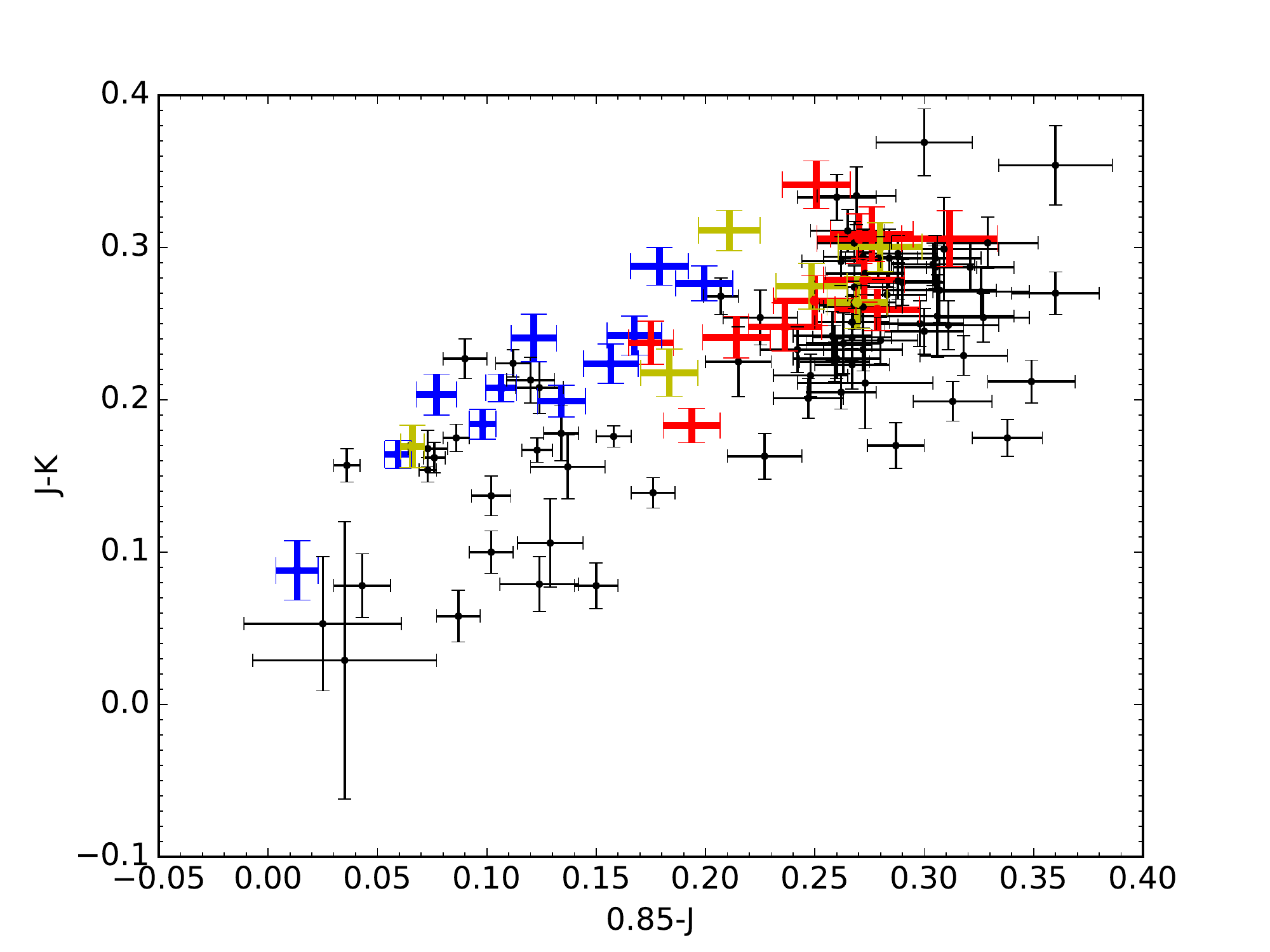}
\end{center}
\caption{Two-color plot derived from the near-infrared spectra of Hildas (colored points) and Trojans (black points). Hildas that are classified as less-red (LR) and red (R) via visible color and/or infrared reflectivity are denoted by blue and red, respectively; unclassifed objects are marked in yellow. The Hilda near-infrared color distribution lacks the robust bimodality evident in the Trojan data. Nevertheless, LR and R Hildas occupy distinct regions in color space, implying systematically different surface properties and allowing us to classify an additional 4 objects into the LR and R sub-populations.} \label{colors}
\end{figure}

Figure~\ref{colors} shows the distribution of Hilda colors (colored points) in the $J-K$ vs. $0.85-J$ space, along with the corresponding distribution of Trojan colors (black points), taken from \citet{emery}. The colors of the LR and R Hildas are denoted by blue and red. The overall Trojan color distribution is robustly bimodal, with the locations of the LR and R Trojans clearly identifiable. In contrast, while the LR and R Hildas in our near-infrared spectral sample occupy disparate regions in the $J-K$ vs. $0.85-J$ space, these regions overlap, resulting in an overall Hilda color distribution that lacks the bimodality seen in the Trojan data. Taking the positions of the 6 hitherto unclassified Hildas on the two-color plot into consideration, we are able classify one (2067 Aksnes) as LR and three (3561 Devine, 5661 Hildebrand, and 14669 Beletic) as R, with the remaining two Hildas unclassifiable due to their having near-infrared color indices that lie in the overlap of the LR and R color regions. The final classification count in our spectral sample is 12 LR and 14 R.

Examining the color indices more closely, we see that the R Trojans and R Hildas have comparable near-infrared colors, in line with the observation from our comparison of their average spectra in Section~\ref{subsec:trojans}. Meanwhile, the LR Hildas have redder (i.e., higher) color indices than the LR Trojans. The relative shift in the cluster of LR Hildas relative to the LR Trojans is particularly noticeable along the $0.85-J$ axis, in agreement with the visual spectroscopic comparison in Figure~\ref{trojans}, which shows that the LR Hildas have steeper spectral slopes than LR Trojans in the short wavelength end of IRTF/SpeX wavelength range.

\subsection{Implications}\label{subsec:implications}

In the context of general characteristics shared by Hildas and Trojans, e.g., reddish colors, near-identical optical albedos, and bimodalities in visible color and infrared reflectivity, the large-scale similarity in near-infrared spectra offers further support to the idea of a common origin for these two minor body populations. While the overall near-infrared color distribution of Hildas does not display bimodality, as in the case of Trojans, Hildas classified as LR or R via visible color and/or infrared reflectivity cluster around two separate near-infrared color centers. 

The systematic spectral differences between LR and R objects within the Hilda and Trojan populations, which are manifested from the visible to the infrared, are indicative of distinct surface compositions \citep[see also discussion in][]{emery}. \citet{wong3} developed a hypothesis within the framework of current dynamical instability models to explain the origin of the sub-populations attested in the Hildas, Trojans, and similarly-sized Kuiper Belt objects \citep{wong5}. In short, the precursor objects to these three populations initially formed in a planetesimal disk spanning a wide range of heliocentric distances in the outer solar system beyond the orbits of the ice giants with roughly cometary compositions rich in volatile ices such as ammonia, methanol, hydrogen sulfide, etc. Subsequent insolation-driven sublimation loss led to the development of distinct surface compositions over the course of the first few hundred million years after the end of planet formation, prior to the onset of dynamical stability that scattered these bodies into their present-day locations.

From our modeling, we predict hydrogen sulfide ice (H$_{2}$S) to be the crucial volatile ice species responsible for the development of two distinct spectral types. Objects in this outer solar system primordial disk would have been split into two groups --- the objects situated closer to the Sun experienced higher surface temperatures and became depleted of H$_{2}$S on their surfaces, while objects farther out retained H$_{2}$S. Previous experimental work has shown that irradiation of volatile ices leads to a general reddening of the optical color \citep[e.g.,][]{brunetto}. We posit that the presence of H$_{2}$S, which is known to induce a strong reddening of the surface upon irradiation \citep[see, for example, H$_{2}$S frost on Io;][]{carlson}, would contribute additional reddening relative to the case where H$_{2}$S was absent. In both cases, the result of irradiation would have been the formation of a refractory mantle on the surface of the objects, with the presence or absence of sulfur-bearing minerals yielding a color bimodality among objects in the primordial planetesimal disk. Objects that retained H$_{2}$S on their surfaces would develop R colors, while those that became depleted in H$_{2}$S would develop LR colors. The subsequent dynamical instability spread these objects across the present-day minor body populations, which therefore inherit this primordial color bimodality.

Our present analysis has shown that LR Hildas have relatively steeper spectral slopes in the near-infrared when compared to LR Trojans; meanwhile, R Hildas and Trojans have near-identical visible and near-infrared colors. In the context of the aforementioned color bimodality hypothesis, which posits that both LR and R Hildas and Trojans were scattered into their present-day orbits from the same progenitor population in the outer Solar System, this disparity between LR Hildas and Trojans is likely explained by their different present-day environments. Hildas orbit significantly closer to the Sun than Trojans and thereby experience higher surface temperatures. It follows that the surface chemistries (sulfur-bearing or sulfur-less) of LR and R objects react differently to heating. While R objects appear to be stable to the temperature change between the Trojan and Hilda regions, LR objects develop redder spectral slopes in the near-infrared and bluer optical colors when heated.

The putative broad and shallow absorption feature in the average LR Hilda spectrum between 0.5 and 1.0~$\mu$m (Section~\ref{subsec:trojans}) is notable in that it suggests the first spectroscopic signature to be found at visible or near-infrared wavelengths on Hildas. Future study of more precise spectra of Hildas with continuous wavelength coverage across the visible and near-infrared is needed to confirm and/or characterize this feature. Nevertheless, comparing the Hildas with other nearby minor body populations offers a clue to its possible origin. 

Looking inward, we find some clearer examples of this feature. Combined visible and near-infrared spectra of two Cybele group asteroids --- 76 Freia and 107 Camilla --- show a pronounced concave-up shape throughout the 0.5--1.0~$\mu$m region within otherwise featureless spectra \citep{takiremery}. These objects lie in the 7:4 mean-motion resonance with Jupiter, with orbital semimajor axes of 3.41 and 3.49~au, respectively, and, similar to Hildas and Trojans, are presumed to originate in the outer solar system within the dynamical instability scenario \citep{roig2}. Both of these Cybele asteroids have LR visible colors and, when considered alongside the LR Hildas and Trojans, point to an intriguing trend: the depth of the broad 0.5--1.0~$\mu$m feature appears to increase with decreasing heliocentric distance.

This broad absorption feature between 0.5 and 1.0~$\mu$m in asteroidal spectra is generally interpreted as indicating hydrated phyllosilicate minerals on the surface \citep[e.g.,][and references within]{rivkin}. The possible presence of hydrated minerals on the surface of Hildas has major implications for their evolutionary history, since it would require melting of surficial water ice, which does not occur at the present-day surface temperatures of Hildas. Within the framework of dynamical instability models, the aforementioned trend between the depth of the 0.7~$\mu$m feature and heliocentric distance among LR objects could be explained if the Cybeles and Hildas were scattered onto orbits that passed systematically closer to the Sun than Trojans during the period of dynamical instability, thereby experiencing higher surface temperatures and possible melting of water ice. The additional observation that R Hildas do not display this absorption signature may indicate that the presence of sulfur-bearing components on the surface blocks the development of hydrated mineral absorption features. Further spectral modeling and dynamical instability orbital simulations are needed to determine the specific chemical species responsible for the absorption feature and how it figures into the origin and evolution of these minor body populations.

\section{Conclusion}
In this paper, we presented new near-infrared spectra of 25 Hilda asteroids. As in the case of analogous spectra of Trojans \citep{emery}, we did not detect any absorption features within the wavelength range covered by the IRTF/SpeX spectrograph ($0.7-2.5$~$\mu$m). Classifying the Hildas into less-red and red sub-populations based on their previously published visible color and/or infrared albedo, we found that the average less-red and red Hilda spectra have very distinct shapes. Taken together, the systematic differences in spectroscopic/photometric properties from the visible through the infrared firmly establish that the two Hilda sub-populations possess distinct intrinsic surface compositions, in agreement with the conclusions of earlier studies of Trojans.

Combining our near-infrared spectra with visible spectra from the literature and comparing the results with analogous data for Trojans, we observed that the corresponding sub-populations within the Trojans and Hildas have similar overall spectral shapes. This lends support to the idea that the less-red and red Trojans and Hildas are each sourced from a single progenitor population, as is posited by recent dynamical instability theories of solar system evolution. Upon closer inspection, we uncovered a notable difference between the less-red and red objects in the Hilda and Trojan populations. Whereas the red Hildas and Trojans have nearly identical spectra, the less-red Hildas are significantly redder than less-red Trojans in the near-infrared, while being somewhat bluer than their Trojan counterparts at visible wavelengths. From this observation, we inferred the presence of a very broad and shallow absorption feature in the average less-red Hilda spectrum between 0.5 and 1.0~$\mu$m, suggesting the possible presence of hydrated minerals on the surface. In the context of our hypothesis regarding the origin of the two sub-populations, we proposed that the discrepancy between less-red Hilda and Trojan spectra is a consequence of their different surface temperatures. Meanwhile, objects in the red spectral type are stable to the temperature gradient between the Hilda and Trojan regions.

Pinning down a more detailed picture of the surface (and bulk) composition of Hildas and Trojans remains a crucial and unfulfilled step in the overarching objective of understanding the origin of these minor body populations and validating current models of solar system evolution. A potentially fruitful avenue of further study is analyzing new Hilda spectra covering the 3~$\mu$m region. Previous study of Trojans has revealed a significant absorption feature centered at around 3.1~$\mu$m, consistent with both water frost and ammonia ice \citep{brown}. Intriguingly, the depth of this feature is strongly correlated with spectral type, with less-red Trojans having systematically deeper absorption than red Trojans. This absorption signal has already been detected in the three extant Hilda spectra covering this wavelength range \citep{takiremery}; however, the three objects all belong to the less-red sub-population. Additional spectroscopic observations targeting both less-red and red Hildas will provide a more detailed look into this feature in relation to the two sub-populations and will aid in the continued exploration of environmental effects on the surface properties of these primordial bodies.

\section*{Appendix}

Plotted here are the final reduced near-infrared spectra of Hildas presented in this paper. Of these, 25 are derived from the IRTF/SpeX observations described in Section~\ref{sec:obs}, and three (153 Hilda, 190 Ismene, and 361 Bononia) are previously published in \citet{takiremery}. All spectra are normalized to unity at 2.2~$\mu$m and shifted in intervals of 0.5 for clarity. The gray bars denote the main wavelength regions of telluric water vapor absorption. Residual signals within these bands are caused by variable atmospheric conditions during and/or between observations of asteroids and solar analogs stars used for telluric correction.

\begin{figure*}[t!]
\begin{center}
\includegraphics[width=16cm]{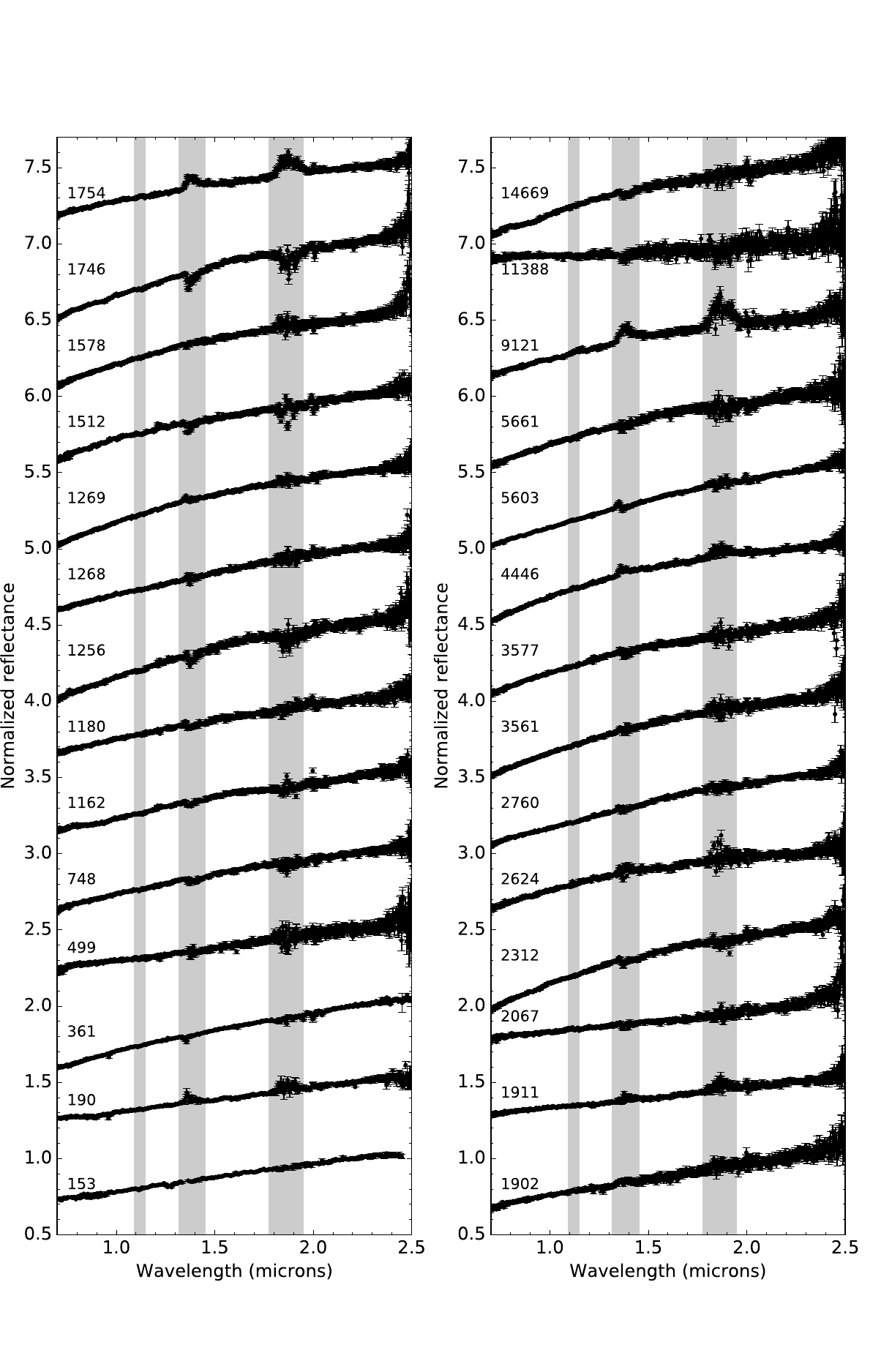}
\end{center}
\label{allspectra}
\end{figure*}


\small

\begin{thebibliography}{}
\providecommand{\natexlab}[1]{#1}
\providecommand{\url}[1]{\texttt{#1}}
\expandafter\ifx\csname urlstyle\endcsname\relax
  \providecommand{\doi}[1]{doi: #1}\else
  \providecommand{\doi}{doi: \begingroup \urlstyle{rm}\Url}\fi
 
 \bibitem[Brown(2016)]{brown}
 Brown, M.~E. 2016, AJ, 152, 159

\bibitem[Brunetto et~al.(2006)Brunetto, Barucci, Dotto, and Strazzulla]{brunetto}
Brunetto, R., Barucci, M.~A., Dotto, E., \& Strazzulla, G. 2006, ApJ, 644, 646

\bibitem[Bus \& Binzel(2002)]{bus}
Bus, S.~J., \& Binzel, R.~P. 2002, Icar, 158, 106

\bibitem[Carlson et~al.(2007)]{carlson}
Carlson, R.~W., Kargel, J.~S., Doute, S., Soderblom, L.~A., \& Brad Dalton, J. 2007, in Io After Galileo, ed. R.~M.~C. Lopes, \& J.~R. Spencer (Chichester, UK: Praxis Publishing Ltd), 194

\bibitem[Cushing et~al.(2004)]{cushing}
Cushing, M.~C., Vacca, W.~D., \& Rayner, J.~T. 2004, PASP, 116, 362

\bibitem[Dahlgren \& Lagerkvist(1995)]{Hildaspectra}
Dahlgren, M., \& Lagerkvist, C.-I. 1995, A\&A, 302, 907

\bibitem[Dahlgren et~al.(1997)]{Hildaspectra2}
Dahlgren, M., Lagerkvist, C.-I., Fitzsimmons, A., Williams, I.~P., \& Gordon, M. 1997, A\&A, 323, 606

\bibitem[Emery \& Brown(2004)]{emerybrown}
Emery, J.~P., \& Brown, R.~H. 2004, Icar, 170, 131

\bibitem[Emery et~al.(2011)Emery, Burr, and Cruikshank]{emery}
Emery, J.~P., Burr, D.~M., \& Cruikshank, D.~P. 2011, AJ, 141, 25

\bibitem[Fern{\' a}ndez et~al.(2009)]{fernandez}
Fern{\' a}ndez, Y.~R., Jewitt, D., \& Ziffer, J.~E. 2009, AJ, 138, 240

\bibitem[Franklin et~al.(2004)]{franklin}
Franklin, F.~A., Lewis, N.~K., Soper, P.~R., \& Holman, M.~J. 2004, AJ, 128, 1391

\bibitem[Gil-Hutton \& Brunini(2008)]{gilhutton}
Gil-Hutton, R., \& Brunini, A. 2008, Icar, 193, 567

\bibitem[Gomes et~al.(2005)Gomes, Levison, Tsiganis, and Morbidelli]{gomes}
Gomes, R., Levison, H.~F., Tsiganis, K., \& Morbidelli, A. 2005, Natur, 435, 466

\bibitem[Grav et~al.(2012)]{grav}
Grav, T., Mainzer, A.~K., Bauer, J.~M., et al. 2012, ApJ, 744, 197

\bibitem[Lazzaro et~al.(2004)]{lazzaro}
Lazzaro, D., Angeli, C.~A., Carvano, J.~M., et al. 2004, Icar, 172, 179

\bibitem[Morbidelli et~al.(2005)Morbidelli, Levison, Tsiganis, and
  Gomes]{morbidelli}
Morbidelli, A., Levison, H.~F., Tsiganis, K., \& Gomes, R. 2005, Natur, 435, 462

\bibitem[Rayner et~al.(2003)]{rayner}
Rayner, J.~T., Toomey, D.~W., Onaka, P.~M., et al. 2003, PASP, 115, 362

\bibitem[Rivkin et~al.(2015)]{rivkin}
Rivkin, A.~S., Campins, H., Emery, J.~P., et~al. 2015, in Asteroids IV, ed. P. Michel, F.~E. DeMeo, \& W.~F. Bottke (Tucson, AZ: University of Arizona Press), 65

\bibitem[Roig \& Nesvorn{\' y}(2015)]{roig2}
Roig, F., \& Nesvorn{\' y}, D. 2015, AJ, 150, 186

\bibitem[Roig et~al.(2008)Roig, Ribeiro, and Gil-Hutton]{roig}
Roig, F., Ribeiro, A.~O., \& Gil-Hutton, R. 2008, A\&A 483, 911

\bibitem[Ryan \& Woodward(2011)]{Hildaalbedo}
Ryan, E.~L., \& Woodward, C.~E. 2011, AJ, 141, 186

\bibitem[Takir \& Emery(2012)]{takiremery}
Takir, D., \& Emery, J.~P. 2012, Icar, 219, 641

\bibitem[Vilas et~al.(1994)]{vilas}
Vilas, F., Jarvis, K.~S., \& Gaffey, M.~J. 1994, Icar, 109, 274

\bibitem[Wong \& Brown (2016)Wong, Brown]{wong3}
Wong, I., \& Brown, M.~E. 2016, AJ, 152, 90

\bibitem[Wong \& Brown (2017a)Wong, Brown]{wong4}
Wong, I., \& Brown, M.~E. 2017a, AJ, 153, 69

\bibitem[Wong \& Brown (2017b)Wong, Brown]{wong5}
Wong, I., \& Brown, M.~E. 2017b, AJ, 153, 145

\bibitem[Wong et~al.(2014)Wong, Brown, and Emery]{wong}
Wong, I., Brown, M.~E., \& Emery, J.~P. 2014, AJ, 148, 112

\bibitem[Yang \& Jewitt(2007)]{yangjewitt}
Yang, B., \& Jewitt, D. 2007, AJ, 134, 223


\end{thebibliography}
\end{document}